\documentclass[aps,prl,twocolumn,superscriptaddress,preprintnumbers,floatfix,10pt,nofootinbib]{revtex4-1}
\pdfoutput=1
\usepackage{hyperref}
\usepackage{amsmath}
\usepackage{amsfonts}
\usepackage{slashed}
\usepackage{amssymb}
\usepackage{bm}
\usepackage{bbold}
\usepackage{graphicx}
\usepackage{enumitem}
\usepackage{xcolor}

\newcommand{\refcite}[1]{Ref.~\cite{#1}}
\newcommand{\refscite}[1]{Refs.~\cite{#1}}

\newcommand{\eq}[1]{Eq.~\eqref{eq:#1}}

\newcommand{\Eqs}[2]{Eqs.~\eqref{eq:#1} and \eqref{eq:#2}}

\newcommand{\fig}[1]{Fig.~\ref{fig:#1}}

\newcommand{\tab}[1]{Table~\ref{tab:#1}}

\newcommand{\nn}{\nonumber}

\newcommand{\abs}[1]{\lvert#1\rvert}

\newcommand{\ord}[1]{\mathcal{O}(#1)}

\newcommand{\exv}[1]{\langle#1\rangle}

\newcommand{\ket}[1]{\lvert#1\rangle}

\newcommand{\Sl}[1]{\slashed{#1}}

\newcommand{\df}{\mathrm{d}}

\newcommand{\img}{\mathrm{i}}

\renewcommand{\vec}[1]{\bm{#1}}

\renewcommand{\tensor}[1]{\textsf{#1}}

\newcommand{\eps}{\epsilon}
\newcommand{\la}{\lambda}

\newcommand{\bq}{{\bar q}}

\newcommand{\bd}{{\bar d}}
\newcommand{\bu}{{\bar u}}
\newcommand{\bs}{{\bar s}}
\newcommand{\bc}{{\bar c}}
\newcommand{\bb}{{\bar b}}

\newcommand{\bQ}{{\bar Q}}

\newcommand{\cL}{\mathcal{L}}

\newcommand{\ty}{{\tilde y}}

\newcommand{\alphas}{\alpha_s}
\newcommand{\Ecm}{E_\mathrm{cm}}
\newcommand{\lqcd}{\Lambda_\mathrm{QCD}}
\newcommand{\as}{\alpha_{s}}
\newcommand{\MSbar}{$\overline{\text{MS}}$}
\newcommand{\SM}{\mathrm{SM}}

\newcommand{\dis}{\mathrm{dis}}
\newcommand{\fav}{\mathrm{fav}}

\newcommand{\stat}{\mathrm{stat}}
\newcommand{\syst}{\mathrm{syst}}

\newcommand{\MeV}{\,\mathrm{MeV}}
\newcommand{\GeV}{\,\mathrm{GeV}}
\newcommand{\TeV}{\,\mathrm{TeV}}

\newcommand{\fb}{\,\mathrm{fb}}
\newcommand{\ab}{\,\mathrm{ab}}

\newcommand{\CL}{\,\mathrm{CL}}

\allowdisplaybreaks[2]

\providecommand{\sectionPaper}[1]{\section{\boldmath #1}}
\providecommand{\headingAcknowledgments}{\vspace{0.4em}\paragraph{Acknowledgments:}}
\providecommand{\whereIsTheSupplement}{\hyperref[supplement]{Supplemental Material}}

\begin{document}

%%%%%%%%%%%%%%%%%%%%%%%%%%%%%%%%%%%%%%%%%%%%%%%%%%%%%%%%%%%%%%%%%%%%%%%%%%%%%%%%
% Title page
%%%%%%%%%%%%%%%%%%%%%%%%%%%%%%%%%%%%%%%%%%%%%%%%%%%%%%%%%%%%%%%%%%%%%%%%%%%%%%%%

\preprint{\vbox{\hbox{Nikhef 2025-010}}}
\title{
Detecting Traces of Light-Quark Yukawa Couplings
\texorpdfstring{\\}{}
to the Higgs Boson in Fragmentation Products
}

\author{Johannes K. L. Michel}%
\email{j.k.l.michel@uva.nl}%
\affiliation{Institute for Theoretical Physics Amsterdam and Delta Institute for Theoretical Physics, University of Amsterdam, Science Park 904, 1098 XH Amsterdam, The Netherlands}%
\affiliation{Nikhef, Theory Group, Science Park 105, 1098 XG, Amsterdam, The Netherlands}%

\date{May 22, 2025}

%%%%%%%%%%%%%%%%%%%%%%%%%%%%%%%%%%%%%%%%%%%%%%%%%%%%%%%%%%%%%%%%%%%%%%%%%%%%%%%%
\begin{abstract}
We point out that Yukawa interactions of light quarks
with the Higgs boson are imprinted
as unique azimuthal modulations in the density of fragmentation hadrons
relative to the Higgs $p_T$.
We introduce Yukawa Fragmentation Asymmetries (YFAs),
interference observables
that are linearly proportional to real (Standard Model) or CP-odd Yukawa couplings, respectively.
The chiral suppression is lifted nonperturbatively
by chiral-odd multi-hadron fragmentation functions.
As a simple example process, we consider $VH$ production
with a tagged target fragmentation hadron at the HL-LHC.
We find promising projected sensitivities
to first and second-generation couplings
while maintaining superior control over theory systematics,
flavor separation, and model independence compared to present methods.
Our results point to deep synergies between precision studies of confinement
and the in-depth exploration of the Higgs sector.

\end{abstract}
%%%%%%%%%%%%%%%%%%%%%%%%%%%%%%%%%%%%%%%%%%%%%%%%%%%%%%%%%%%%%%%%%%%%%%%%%%%%%%%%

\maketitle

%%%%%%%%%%%%%%%%%%%%%%%%%%%%%%%%%%%%%%%%%%%%%%%%%%%%%%%%%%%%%%%%%%%%%%%%%%%%%%%%
\sectionPaper{Introduction}
\label{sec:intro}
%%%%%%%%%%%%%%%%%%%%%%%%%%%%%%%%%%%%%%%%%%%%%%%%%%%%%%%%%%%%%%%%%%%%%%%%%%%%%%%%

\begin{figure}[b]
\includegraphics[width=0.5\textwidth]{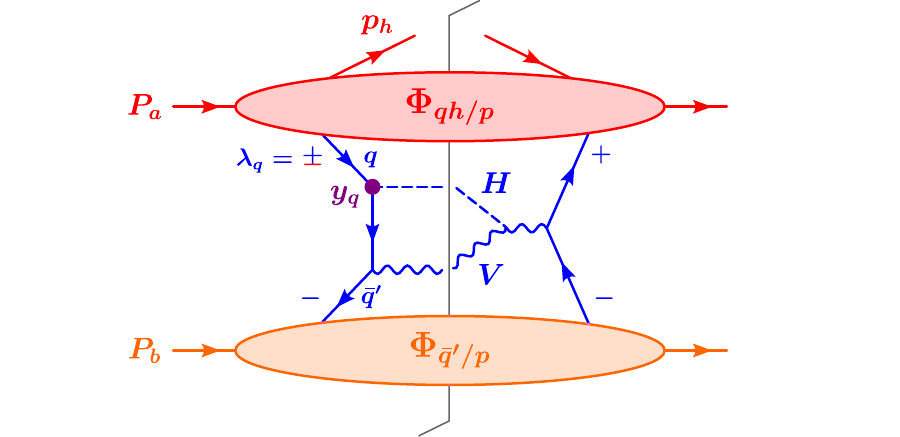}
\caption{
Factorization for $pp$ collisions with a tagged target fragmentation hadron.
As an example of a hard scattering diagram (blue),
we illustrate the interference of the Yukawa interaction with standard SM $VH$ production
and indicate representative helicity configurations.
\label{fig:diagram_pp_VH}
}
\end{figure}

Thirteen years after the discovery of the Higgs boson~\cite{ATLAS:2012yve, CMS:2012qbp},
evidence for many of its predicted couplings to leptons and quarks
remains elusive.
While the most exotic fermions in the Standard Model (SM) of particle physics
were the first to have their Yukawa interactions
with the Higgs boson confirmed~\cite{CMS:2022dwd, ATLAS:2022vkf},
owing precisely to the large masses (and thus large couplings)
that made them hard to discover in the first place,
it is a great irony
that such an observation is extremely challenging to make
for some of the lightest, most mundane, and most abundant matter particles,
the up and down quarks
that form the protons and neutrons of everyday life~\cite{Salam:2022izo}.
Curiously, within the first fermion generation
the up quark, with charge $Q_u = 2/3$, is significantly \emph{lighter}
than the down quark with $Q_d = -1/3$~\cite{Tanabashi:2018oca}.
If the roles were reversed,
as in the other two generations,
or if the mass difference were too small to counter
the stronger electromagnetic repulsion in the $(uud)$ proton~\cite{BMW:2014pzb},
the $(ddu)$ neutron would be the lightest baryon,
making chemistry impossible, and thus Life.
Empirically confirming that this curious fact is indeed rooted
in the relative size of the $u$ and $d$ Yukawa couplings to the Higgs field,
as theorized in the SM,
remains a pressing unsolved problem for 21\textsuperscript{st}-century particle physics.

A variety of experimental probes of first and second-generation Yukawa
couplings at the upcoming high-luminosity (HL) LHC
have been proposed, ranging
from exclusive~\cite{Bodwin:2013gca, Kagan:2014ila, Konig:2015qat, dEnterria:2025rjj}
and inclusive decays~\cite{Perez:2015aoa, Perez:2015lra, Carpenter:2016mwd},
to off-shell rates~\cite{Zhou:2015wra, Falkowski:2020znk, Vignaroli:2022fqh, Balzani:2023jas}
and differential distributions
in (associated) Higgs production~\cite{Brivio:2015fxa, Bishara:2016jga, Soreq:2016rae, Yu:2016rvv, Alasfar:2019pmn, Aguilar-Saavedra:2020rgo, Alasfar:2022vqw}.
Nonetheless, experimental measurements to date have
only been able to set the expected upper
limits~\cite{ATLAS:2017gko, ATLAS:2018xfc, CMS:2019hve, ATLAS:2022ers, CMS:2022fsq, ATLAS:2022fnp, CMS:2023rcv, ATLAS:2024ext, CMS:2024tgj, CMS:2025xkn},
and even HL-LHC projections
range
from tantalizingly close to a long way off from discovery:
At $95\%\CL$, one expects to be able to indirectly constrain~\cite{deBlas:2019rxi}
%%%
\begin{align}
\abs{y_q/y_q^\SM} < 560\,,~260\,,~13\,,~1.2
\,,\end{align}
%%%
for $q = u, d, s, c$ from the Higgs boson total width.
This would leave much room for nonstandard, beyond-the-SM (BSM)
Yukawa interactions~\cite{Erdelyi:2024sls},
and calls for innovative and complementary approaches
to strengthen the projected limits
at the HL-LHC and future colliders~\cite{Gao:2016jcm, Duarte-Campderros:2018ouv, Li:2019xwd, Bi:2020frc, Knobbe:2023njd, Kamenik:2023hvi, Liang:2023wpt}.
Experimental searches for light-quark Yukawa interactions
face three key challenges:
\textbf{(1)}~At its most basic, they contend
with the smallness of the coupling,
making for tiny signal rates.
\textbf{(2)}~In all known cases, one faces
irreducible backgrounds from other SM processes
and the Yukawa couplings of heavier quarks,
which require complicated perturbative calculations.
\textbf{(3)}~While light quarks interact perturbatively with the Higgs boson
at high energies, they originate from, and fragment into,
strongly bound hadronic states, requiring rigorous
factorization statements to separate the Yukawa signal
from the nonperturbative physics of confinement.

In this letter we propose the measurement
of \textbf{Yukawa Fragmentation Asymmetries (YFAs)},
a novel class of observables that resolve all three challenges
in one stroke.
They are interference observables
that are \textbf{(1)}~\emph{linearly} proportional
to the Yukawa couplings of interest,
in particular those of the abundant valence $u$ and $d$ quarks in protons and pions,
which improves sensitivity.
They \textbf{(2)} feature powerful, symmetry-protected mechanisms
that stabilize them against radiative corrections
and the contributions of heavier sea quarks.
Most notably, they turn challenge~\textbf{(3)} into a virtue
and --- within the factorization paradigm --- make use
of the nontrivial properties of the QCD confinement transition itself
to overcome \textbf{(1)} and \textbf{(2)}.

The remainder of this letter is structured as follows:
We first present an accessible introduction
to the physical mechanism behind YFAs,
focusing on the simple example of $VH$ production
with a tagged target fragmentation hadron $h$ at the LHC.
We next list some of the distinguishing all-order properties of YFAs.
(Derivations and additional field-theoretical background
are presented in a companion paper~\cite{paper2}.)
We then present sensitivity estimates for YFAs in $VHh$ production at the HL-LHC
for first and second-generation quarks.
We close by listing applications to other processes and at future colliders.

%%%%%%%%%%%%%%%%%%%%%%%%%%%%%%%%%%%%%%%%%%%%%%%%%%%%%%%%%%%%%%%%%%%%%%%%%%%%%%%%
\sectionPaper{Physical mechanism}
\label{sec:physical_mechanism}
%%%%%%%%%%%%%%%%%%%%%%%%%%%%%%%%%%%%%%%%%%%%%%%%%%%%%%%%%%%%%%%%%%%%%%%%%%%%%%%%

We are interested in constraining the parameters
of the following effective Lagrangian
coupling the Higgs boson $H$ to quark fields $\psi_q$~\cite{Konig:2015qat},
%%%
\begin{align} \label{eq:lagrangian}
\cL \supset
- \frac{y_q}{\sqrt{2}} H \bar{\psi}_q \psi_q
- \frac{\img \ty_q}{\sqrt{2}} H \bar{\psi}_q \gamma_5 \psi_q
\,.\end{align}
%%%
Within the SM, $y_q^\SM = \sqrt{2} \, m_q/v$,
where $m_q$ is the quark mass and $v$ is the Higgs vacuum expectation value,
while CP-odd interaction terms vanish, $\ty_q^\SM = 0$.

A key challenge in constructing interference observables
linear in the quark Yukawa couplings $y_q$ and $\ty_q$
is that interference diagrams with other SM processes
(as shown in blue in \fig{diagram_pp_VH})
are \emph{chirally suppressed} for massless quarks in unpolarized partonic collisions.
In standard $VH$ production with $V = Z, W^\pm$
as shown to the right of the cut,
the helicities of the quark and antiquark must be opposite
because an SM interaction term like
$\bar{\psi}_{q'} \gamma^\mu \psi_{q}
= \bar{\psi}_{q'L} \gamma^\mu \psi_{qL}
+ \bar{\psi}_{q'R} \gamma^\mu \psi_{qR}$
(which annihilates these helicity states)
only couples field components of identical handedness to each other,
i.e., it is \emph{chiral even}.
The situation is different on the left
because \eq{lagrangian} is \emph{chiral odd}
and couples left to right-handed fields,
thus a nonzero amplitude for a negative-helicity antiquark
requires quark helicity $\la_q = -$ (red).
However, in unpolarized collisions one only sums
over identical helicity configurations in both the amplitude
and the conjugate amplitude.
Getting back $\la_q = +$ (blue)
requires an additional insertion of a mass term $m_q \bar{\psi}_q \psi_q$
and leads to an overall suppression by $y_q \, m_q/Q \sim y_q^2$,
with $Q^2 = (p_V + p_H)^2$ the hard scale,
on par with the Yukawa amplitude squared.

One important insight is that the helicity configuration where
one interferes $\la_q = -$ with $+$
is, in fact, physical, and corresponds to \emph{transverse} quark polarization.
Indeed, in a basis of helicity eigenstates $\ket{\pm}$,
the spin density matrix of a transversely polarized quark
involves a linear combination
of offdiagonal Pauli matrices $\sigma_1$ and $\sigma_2$.
If the incoming fermion can be prepared in such a state,
the chiral suppression of the interference diagram in \fig{diagram_pp_VH}
is lifted.
The key question is whether
a nonzero transverse Bloch vector
is physically \emph{allowed}.

This is most clearly the case for incoming electrons,
where recently a proposal was made~\cite{Boughezal:2024yjk}
to tune positron beams and transversely polarized
electron beams to the Higgs resonance
to improve the sensitivity to the electron Yukawa coupling
compared to the unpolarized rate~\cite{dEnterria:2021xij}.
This proposal can be adapted to light quarks
by giving one of the incoming nucleons a net transverse polarization,
which would then enter the partonic collision
modulated by the quark transversity parton distribution function (PDF).
This, of course, comes at the significant cost of having to polarize the beam
while maintaining high energies and luminosities.

\begin{figure}
\includegraphics[width=0.5\textwidth]{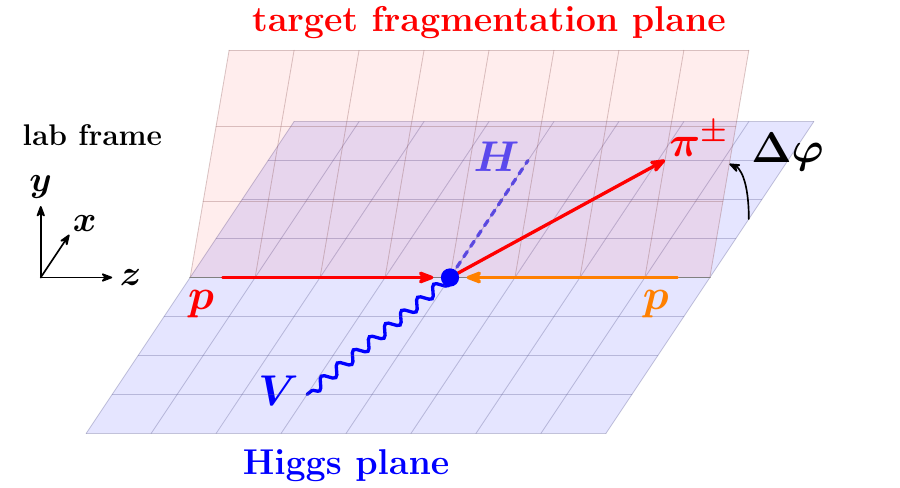}
\caption{
Kinematics of $VH$ production in $pp$ collisions
with a tagged target fragmentation hadron in the lab frame (momenta not to scale).
The YFA for CP-even, SM-like Yukawa interactions
is formed by comparing rates of hadrons above ($\sin \Delta \varphi > 0$)
and below the Higgs plane.
\label{fig:kinematics}
}
\end{figure}

We instead propose a different route that is viable
also in unpolarized hadronic collisions,
and is based on the powerful
theoretical observation~\cite{Collins:1992kk, Artru:1995zu, Collins:1993kq, Boer:1997nt, Boer:2003ya, Sivers:2008my, Sivers:2009qj}
that transverse parton polarization is nonperturbatively
imprinted on the distribution of fragmentation products
in the transverse plane,
leading to e.g.\ the experimentally established Collins~\cite{Belle:2005dmx, Belle:2008fdv, BaBar:2013jdt, BaBar:2015mcn, Belle:2019nve}
and Collins-Artru effects~\cite{Belle:2011cur}.
For our example, assume that we tag
on a final-state hadron $h$
that is observed at forward rapidity
and small transverse momentum $p_T^h \sim \lqcd \ll Q$,
and thus at leading power originates from the remnant
of the struck proton moving in the forward direction,
as illustrated in \fig{kinematics}.
In this case, the hadron carries information
about the ``quark hole'' left by the hard-scattering quark,
whose opposite transverse polarization in particular breaks the azimuthal symmetry
of the proton remnant around the beam axis.
This means that a linear correlation between transverse quark spin and the direction of $\vec{p}_T^h$
becomes allowed at leading twist.
The correlation strength is known
as the quark Boer-Mulders fracture function (BMFrF) $h^\perp_{qh}$~\cite{Boer:1997nt, Sivers:2008my, Sivers:2009qj},%
\footnote{
For (extended) fracture functions in general,
see \refscite{Collins:1350496, Trentadue:1993ka, Berera:1995fj, Grazzini:1997ih, Collins:1997sr, Anselmino:2011ss, Chai:2019ykk, Chen:2021vby}.
}
and is a \emph{chiral-odd} nonperturbative matrix element related to chiral symmetry breaking by the QCD vacuum.
Formally, it appears as the coefficient of a Dirac structure $\img \Sl{p}_T^h \Sl{p}_q$
in the decomposition of the FrF correlator $\Phi_{qh/p}$ in \fig{diagram_pp_VH},
which becomes allowed in addition to the unpolarized term $\Sl{p}_q$ after breaking the azimuthal symmetry.

Returning to \fig{diagram_pp_VH}, we thus expect
to observe, at the level of the hadronic cross section,
modulations in hadron yield
that based on Lorentz invariance
are either proportional to $\vec{p}_T^H \cdot \vec{p}_T^h \propto \cos \Delta \varphi$
or the contraction $\eps(P_{a}, P_{b}, p_H, p_h) \propto \sin \Delta \varphi$
with the Levi-Civita symbol,
where $\Delta \varphi \equiv \varphi_h - \varphi_H$ is the signed azimuthal separation between
$H$ and $h$, see \fig{kinematics}.
The maximal violation of parity by the weak interaction
makes the latter the dominant one multiplying CP-even (real) Yukawa couplings $y_q$.
This modulation leads to a difference between hadron rates $N^+_{Xh}$
observed at $p_{y}^{h} > 0$ compared to the rate $N^-_{Xh}$ at negative $p_y^h < 0$
in a righthanded coordinate system chosen as in \fig{kinematics}.
We therefore propose to measure the asymmetry observable
%%%
\begin{align} \label{eq:def_YFA_hV}
\mathrm{YFA}_{Xh}
\equiv \frac{N^+_{Xh} - N^-_{Xh}}{N^+_{Xh} + N^-_{Xh}}
= \frac{\eps_{X/V\! H} \, \eps_h \, \Delta \sigma_{V\! H h}}{\sigma_{X h}}
\,,\end{align}
%%%
where $X$ is a reconstructed state
that the $V H$ system decays into
at a relative rate $\eps_{X/VH}$,
including experimental efficiencies.
Here $\sigma_{X h}$ is the cross section
for producing $X$ and a tagged
hadron in a forward acceptance volume,
which includes hard QCD background processes
and unsuppressed pile-up, i.e.,
unrelated secondary hadron collisions in the detector.
(Accordingly, $\eps_h$ is the signal efficiency
of pile-up suppression cuts for hadron type $h$.)
The on-shell signal asymmetry cross section in the numerator
is predicted to be
%%%
\begin{align} \label{eq:fact_Delta_sigma_hV}
\Delta \sigma_{V\!Hh}
&= \frac{e^3}{4 \pi \Ecm^2} \sum_{q,q'} \int \! \! \df \Phi \,
C_{qq'V}^{TU} \Bigl(
      y_q \bar{h}^\perp_{qh} f_{\bq'}
      - y_{q'} \bar{h}^\perp_{\bq' h} f_{q}
   \Bigr)
\end{align}
%%%
in terms of a perturbatively calculable coefficient $C^{TU}_{qq'V}$~\cite{supplement},
the (anti)quark BMFrFs $\bar{h}_{ih}^\perp(x_a)$
integrated over the acceptance volume,
standard PDFs $f_j(x_b)$,
and the Yukawa coupling of the respective transversely polarized parton.

%%%%%%%%%%%%%%%%%%%%%%%%%%%%%%%%%%%%%%%%%%%%%%%%%%%%%%%%%%%%%%%%%%%%%%%%%%%%%%%%
\sectionPaper{Key properties}
\label{sec:key_properties}
%%%%%%%%%%%%%%%%%%%%%%%%%%%%%%%%%%%%%%%%%%%%%%%%%%%%%%%%%%%%%%%%%%%%%%%%%%%%%%%%

\Eqs{def_YFA_hV}{fact_Delta_sigma_hV}, apart from their great experimental simplicity,
have several field-theoretic properties~\cite{paper2}
that make them particularly powerful probes of light-quark Yukawa couplings:
\begin{enumerate}[label=\textbf{(\Alph*)}]
   \item
   Forming an asymmetry linear in $p_T^h$
   at the nonperturbative scale, which is an intrinsically chiral-odd, fermionic phenomenon,
   \emph{uniquely} picks out a single chiral-odd contribution at the hard scale
   to all orders in perturbation theory and at leading power in $\lqcd/Q$.
   When reconstructing the Higgs boson near its mass shell,
   the only such term even beyond the SM is \eq{lagrangian}.
   \item
   The quark and antiquark terms feature a relative sign.
   This property is phenomenologically critical,
   since it means that sea quark contributions can cancel,
   leaving behind the valence quark contributions of interest.
   It is likewise stable to all orders,
   and can be understood by noting that the SM Lagrangian (approximately)
   preserves CP, while the $\sin \Delta \varphi$ modulation is P odd.
   \item
   Forming an asymmetry $\mathrm{\tilde{Y}FA}_{h,V}$
   with respect to $p_h^x$ instead,
   one arrives at a signal prediction of the same form as \eq{fact_Delta_sigma_hV}
   and with identical coefficient $C^{TU}_{qq'V}$,
   but with the $y_q$ replaced by $\ty_q$.
   Therefore all of our sensitivity estimates for $y_q/y_q^\SM$
   in this letter
   immediately carry over
   to possible CP-odd $\ty_q$ in units of $y_q^\SM$.
   \item
   While the BMFrF is an \emph{a priori} unknown nonperturbative matrix element,
   it can be extracted using azimuthal correlations
   between \emph{two} target fragmentation hadrons,
   one in each beam direction, with chiral-even hard probes.
   These include high-statistics SM baseline processes like Drell-Yan production.
   In light of the present work,
   we strongly encourage our experimental colleagues
   to initiate differential measurements
   of such azimuthal hadron-hadron correlations.
\end{enumerate}

%%%%%%%%%%%%%%%%%%%%%%%%%%%%%%%%%%%%%%%%%%%%%%%%%%%%%%%%%%%%%%%%%%%%%%%%%%%%%%%%
\sectionPaper{Sensitivity estimates}
\label{sec:sensitivity_estimates}
%%%%%%%%%%%%%%%%%%%%%%%%%%%%%%%%%%%%%%%%%%%%%%%%%%%%%%%%%%%%%%%%%%%%%%%%%%%%%%%%

\begin{figure*}
\includegraphics[width=0.5\textwidth]{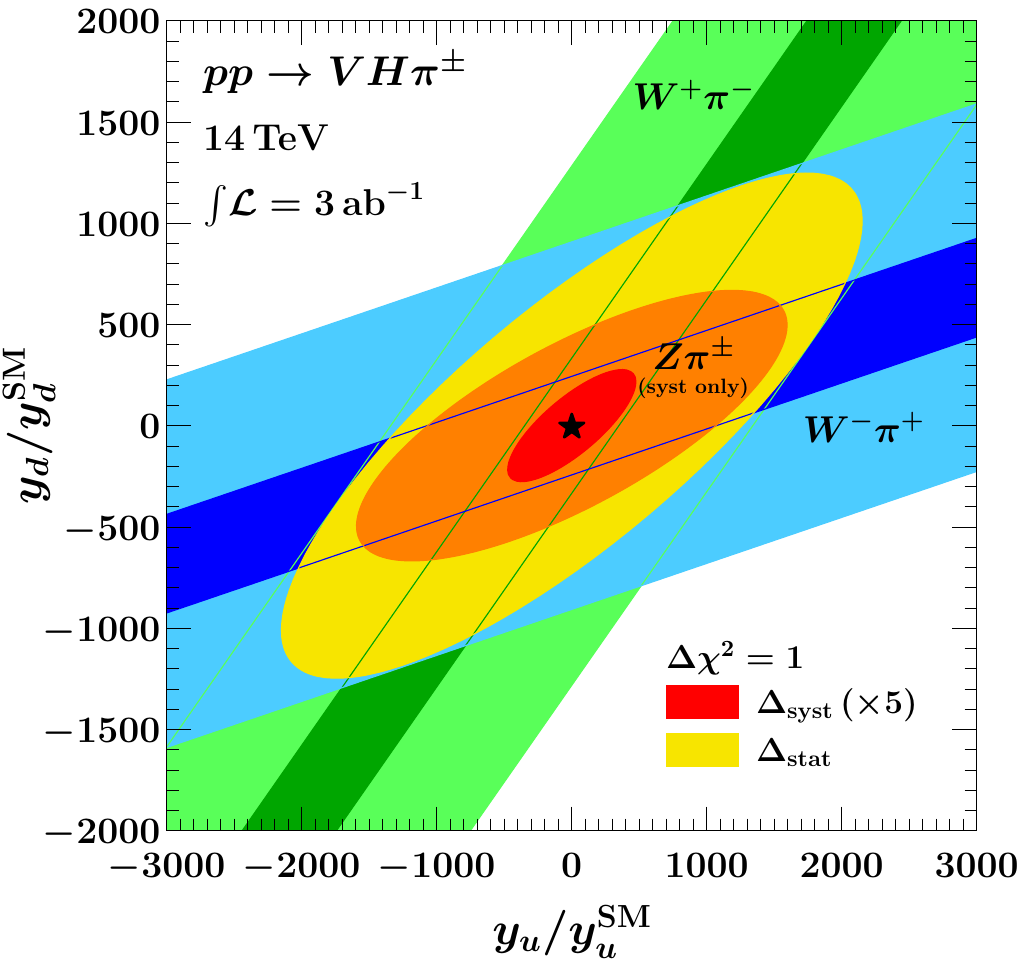}%
\hfill%
\includegraphics[width=0.5\textwidth]{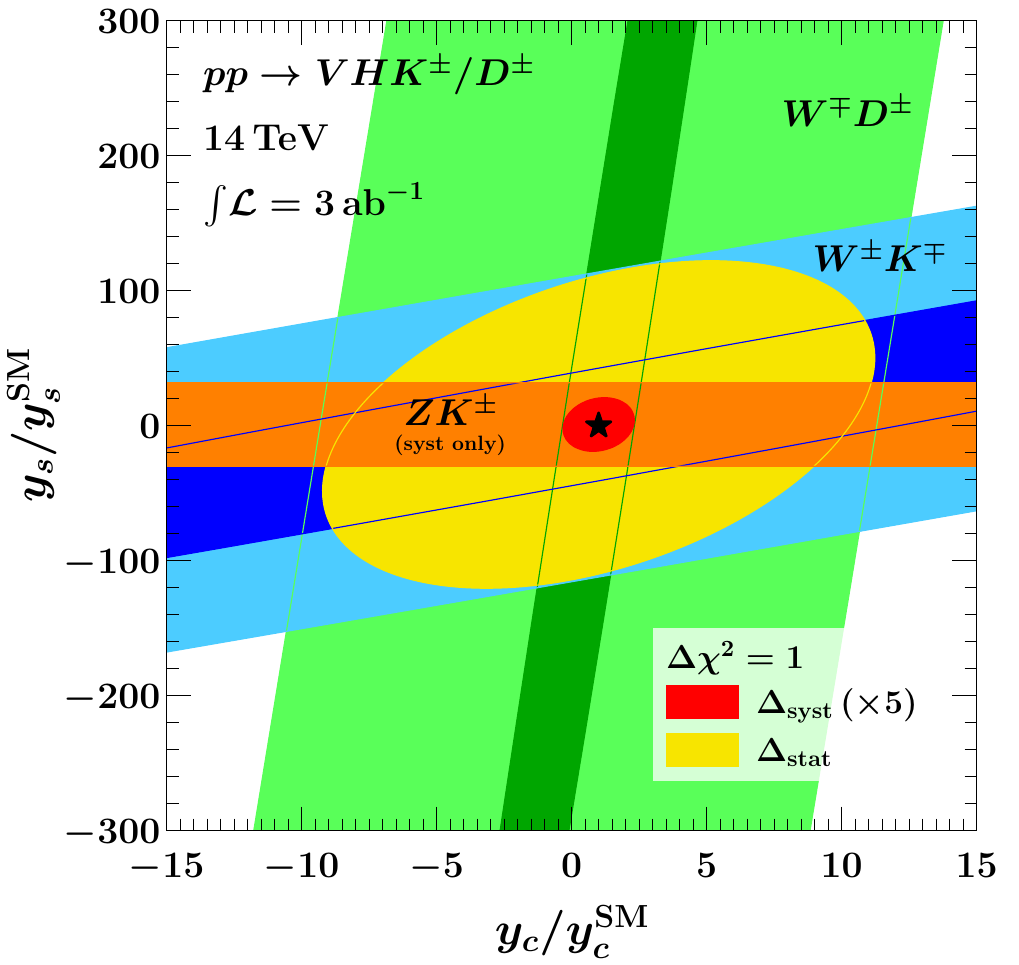}%
\caption{
Projected $\Delta \chi^2 = 1$ contours
from YFA measurements in $VHh$ production at the HL-LHC.
Projected theory uncertainties are scaled up by a factor of five for visibility.
Statistical uncertainties for the $Z$ boson channels
are not competitive with the $W^\pm$ ones and are not shown.
\emph{Left:} Constraints on up and down Yukawa couplings from $h = \pi^\pm$.
\emph{Right:} Constraints on charm and strange Yukawa couplings from $h = K^\pm, D^\pm$.
\label{fig:projected_limits}
}
\end{figure*}

We now derive sensitivity estimates for YFAs in $VH$ production
at the HL-LHC with an integrated luminosity of $3 \ab^{-1}$
at $\Ecm = 14 \TeV$~\cite{ZurbanoFernandez:2020cco}.
We begin with the impact of charged-pion YFAs on $y_u$ and $y_d$.
We envision a scenario where valence BMFrFs have been measured
to sufficient accuracy from baseline processes,
while the smaller sea BMFrFs of heavier (anti)quarks
are still poorly known (since they are not enhanced in the baseline).
The latter then become one of the two main sources of systematic theoretical uncertainty,
together with subleading-twist effects~\cite{supplement}.
We focus on the statistical component of the experimental uncertainty,
leaving experimental systematics to future work.

To evaluate the YFAs, we simply assume here that
%%%
\begin{align} \label{eq:modelling_assumption}
\bar{h}^\perp_{ih}(x) = P_{ih} \, f_{i}(x)
\,,\end{align}
%%%
where $P_{ih}$ is an effective degree of polarization (DOP)
that we take to be independent of the partonic momentum fraction $x$.%
\footnote{Any actual $x$ dependence revealed
by baseline data can of course be included,
and possibly even exploited to enrich the signal.}
Using \eq{modelling_assumption}, the coefficients
multiplying $y_q P_{qh}$ or $y_{q'} P_{\bq'h}$
in \eq{fact_Delta_sigma_hV}
are readily evaluated in terms of standard PDFs~\cite{supplement}.
We also need to estimate $\sigma_{Xh}$ in the denominator of \eq{def_YFA_hV},
which involves the product of the total $X$ background
production cross section $\sigma_X$
and the average yield $\exv{n_h}$ of hadrons of type $h$
produced from the hard scattering.
Using \texttt{Pythia 8.3.16}~\cite{Bierlich:2022pfr} with the default tune,
we find $\exv{n_{\pi^+}} \approx 8.79$ and $\exv{n_{\pi^-}} \approx 8.73$
for signal $VH$ events
within an acceptance of $3 \leq \eta_h \leq 5$ and $0.2 \leq p_T^h \leq 2 \GeV$.%
\footnote{These results are independent
of the vector boson type, and more generally the hard process,
to good approximation.
A more refined treatment
of the background cross section
in terms of integrated unpolarized FrFs $\bar{f}_{qh}(x)$
is again possible with baseline data.}
This results in the following projected statistical uncertainty
on $\Delta \sigma_{V\!Hh}$ in a given channel:
%%%
\begin{align} \label{eq:delta_stat_delta_sigma_VHh}
   \Delta_\stat \bigl( \Delta \sigma_{V\!Hh} \bigr)
=
\frac{1}{\eps_{X/VH}} \sqrt{\frac{
   \exv{n_h} \, \sigma_X / \cL
}{
   2 \, \eps_h \, P^h_\text{PU}
}}
\,,\end{align}
%%%
where $P^{h}_\text{PU}$ is the purity of the hadron sample after pile-up suppression~\cite{supplement},
and we included
a factor of $1/\sqrt{2}$ for the two possible beam directions
in which $h$ may be reconstructed.

A valuable principle to estimate the relative sizes of multi-hadron fragmentation matrix elements
is the degree to which they are \emph{favored} in terms of valence flavor~\cite{Artru:1995zu}.
Applying this principle, together with a suppression by $f_\pi/m_p$ for the most favored contributions
(as appropriate for a chiral-odd matrix element),
we can estimate all DOPs entering \eq{modelling_assumption} for $h = \pi^\pm$~\cite{supplement}.
For $q = s,c,b$ we assign separate, conservative uncertainties for the symmetrized and sea asymmetry DOPs.
For $q = c,b$ we further work to leading power in $\lqcd/m_q$ to sharpen our estimates,
using techniques from \refcite{vonKuk:2023jfd}.

The projected $\Delta \chi^2 = 1$ contours in the $(y_u, y_d)$ plane
that arise from the above statistical and systematic theory uncertainties
are shown in the left panel of \fig{projected_limits}.
We account for the full systematic covariance matrix of experimental channels
that use the same final-state hadron
(and thus are subject to the same unknown sea-quark BMFrFs and sea asymmetries).
If we let $\Delta \kappa_q \equiv y_q/y_q^\SM - 1$
and convert to $95\%\CL$
(constraining one individual parameter at a time),
the projected limits are
%%%
\begin{align}
\frac{\abs{\Delta \kappa_u}}{100}
<
25_\stat
\oplus
1.2_\syst
\,, ~
\frac{\abs{\Delta \kappa_u}}{100}
<
14_\stat
\oplus
0.7_\syst
\,.\end{align}
%%%
We observe that even under our extreme assumptions
on unknown sea-quark BMFrFs,
the inherent cancellation of heavy-quark contributions
in \eq{fact_Delta_sigma_hV} is active
and leads to a subdominant systematic uncertainty.
This underlines a key feature of the observable
apparent from \fig{projected_limits}, namely
that combining different final-state hadrons and bosons
provides a unique and model-independent
handle to tease apart the individual Yukawa couplings.

We also consider the prospects of constraining $y_c$ and $y_s$
from $VH$ YFAs
involving reconstructed kaons $K^\pm$ and charm mesons $D^\pm$.%
\footnote{
We generically use $K^\pm$ ($D^\pm$)
for all hadrons with strangeness $S = \pm 1$ (charm $C = \pm 1$).
}
The relevant yields are
$\exv{n_{K^\pm}} \approx 2.75$
and
$\exv{n_{D^\pm}} \approx \as(m_c)/(2\pi)$,
where the latter is also well in line with the result from \texttt{Pythia 8.3.16}.
The resulting $\Delta \chi^2 = 1$ contours
are shown in the right panel of \fig{projected_limits}.
Converting to $95\%\CL$, we find
%%%
\begin{align}
\abs{\Delta \kappa_c}
< 18_\stat
\oplus
0.5_\syst
\,, ~
\frac{\abs{\Delta \kappa_s}}{10}
<
22_\stat
\oplus
0.8_\syst
\,.\end{align}
%%%

%%%%%%%%%%%%%%%%%%%%%%%%%%%%%%%%%%%%%%%%%%%%%%%%%%%%%%%%%%%%%%%%%%%%%%%%%%%%%%%%
\sectionPaper{Generalizations and extensions}
\label{sec:generalizations_and_extensions}
%%%%%%%%%%%%%%%%%%%%%%%%%%%%%%%%%%%%%%%%%%%%%%%%%%%%%%%%%%%%%%%%%%%%%%%%%%%%%%%%

\begin{figure}
\includegraphics[width=0.5\textwidth]{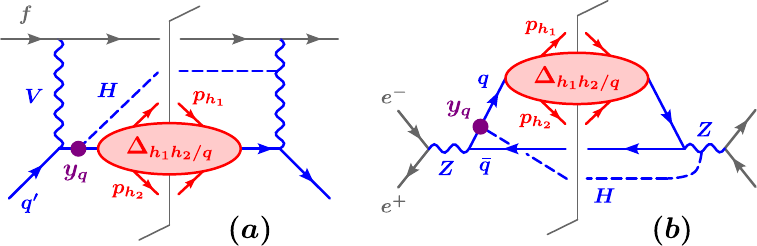}
\caption{
Examples of YFAs that leverage dihadron fragmentation
at the HL-LHC and future colliders.
We suppress PDF correlators for unpolarized partons for simplicity.
\label{fig:other_processes}
}
\end{figure}

Beyond the basic example of $pp \to VHh$ that we chose
here to keep things simple and analytically tractable,
unsuppressed YFAs featuring sea-quark cancellations
can be constructed for any Higgs production process
with a source of parity violation at present and future colliders.
A straightforward generalization is to tag target fragmentation hadrons
in other such processes, including vector boson fusion (VBF),
with much higher rates than $VH$.
A more powerful extension is to exploit the same physical mechanism
in dihadron fragmentation $q, \bar{q} \to h_1 h_2 X$,
where a closely related chiral-odd correlation $H_{h_1 h_2/q}^\sphericalangle$ exists between
the (anti)quark transverse polarization and the relative transverse momentum $R_T^\mu = p_T^{h_1} - p_T^{h_2} \sim \lqcd$
of the two hadrons with respect to the fragmentation axis~\cite{Collins:1993kq, Ji:1993vw, Jaffe:1997hf, Bianconi:1999cd, Bianconi:1999uc, Radici:2001na, Boer:2003ya, Metz:2016swz, Ebert:2021jhy,
Pitonyak:2023gjx, Rogers:2024nhb, Pitonyak:2025lin}.
This \emph{interference dihadron fragmentation function} (IFF)
has already been determined from data~\cite{Courtoy:2012ry, Cocuzza:2023vqs};
a further advantage is that these pairs of nearby hadrons
can typically be reconstructed in the central detector volume.
The IFF induces a YFA in weak Higgs production processes
featuring final-state quarks or antiquarks,
examples of which are shown in \fig{other_processes}.
They range from \textbf{(a)}~VBF at the LHC ($f = q, \bq$),
to the cleaner environment and high rates of VBF at a possible LHeC ($f = e^-$)~\cite{Ahmadova:2025vzd},
and to the pristine conditions of \textbf{(b)}~$VH$ production
with a hadronic $V$ decay at a future $e^+e^-$ collider.
For real, SM-like Yukawa couplings, the relevant azimuthal modulations
are proportional to \textbf{(a)}~$\eps(P, k_f, p_H, R_T)$,
with $P$ the total dihadron momentum
and $k_f$ the momentum of the second jet (or final-state lepton),
and \textbf{(b)}~$\eps(P, q, p_H, R_T)$ with $q^\mu = (\Ecm, \vec{0})$
in the center-of-mass frame.
We leave phenomenological studies of this rich
new field of Higgs precision physics to future work.
We point out that one can also recast our results
in the language of Nucleon Energy and Energy-Energy Correlators~\cite{Cao:2023oef, Guo:2024jch, Chen:2024bpj, Lee:2025okn, Guo:2025zwb, Chang:2025kgq, Kang:2025zto, Herrmann:2025fqy},
use reconstructed forward jets instead of hadrons,
explore structurally similar~\cite{Bacchetta:2023njc} (but Sudakov-suppressed) observables involving transverse momentum-dependent PDFs and FFs,
or extend them into novel search strategies for (pseudo)scalar BSM states.

%%%%%%%%%%%%%%%%%%%%%%%%%%%%%%%%%%%%%%%%%%%%%%%%%%%%%%%%%%%%%%%%%%%%%%%%%%%%%%%%
\sectionPaper{Conclusions}
\label{sec:conclusions}
%%%%%%%%%%%%%%%%%%%%%%%%%%%%%%%%%%%%%%%%%%%%%%%%%%%%%%%%%%%%%%%%%%%%%%%%%%%%%%%%

We have introduced Yukawa Fragmentation Asymmetries (YFAs),
a new class of observables that combine linear sensitivity,
theoretical control, and experimental simplicity
into a uniquely powerful probe of light-quark Yukawa couplings to the Higgs boson.
The observable pulls out all the stops of the Standard Model
to approach these challenging measurement targets,
drawing on confinement, chiral symmetry breaking,
the valence/sea structure of hadrons, and weak parity violation.
In this way, we are able to lift the typical suppression of Yukawa signals
from $y_q^2 \sim y_q \, m_q/v$ to $y_q \, f_\pi/m_p$.
As part of recent renewed interest in interdisciplinary applications
of multi-hadron fragmentation functions~\cite{Bacchetta:2023njc, Guo:2024jch, Chen:2024bpj, Cheng:2025cuv, vonKuk:2025kbv, Lee:2025okn, Guo:2025zwb, Chang:2025kgq, Kang:2025zto, Herrmann:2025fqy},
our results strongly motivate
experimentally exploring these nonperturbative matrix elements~\cite{HERMES:2008mcr, Belle:2011cur, Courtoy:2012ry, STAR:2015jkc, Belle:2017rwm, STAR:2017wsi, AbdulKhalek:2021gbh, COMPASS:2023cgk, Cocuzza:2023vqs},
estimating them from models~\cite{Schweitzer:2012hh},
studying their higher-twist counterparts~\cite{Chen:2023wsi},
and rigorously analyzing them in effective field theories of QCD~\cite{Chen:2001nb, vonKuk:2023jfd, Dai:2023rvd, vonKuk:2024uxe, Copeland:2024cgq, Copeland:2024cgq}.
We expect that our results can become a new cornerstone
of a nascent field~\cite{Furletova:2021wyq, Boughezal:2022pmb, Zhang:2022zuz, Boughezal:2023ooo, Wen:2023xxc, Wen:2024cfu, Wen:2024nff, Curtin:2025ngf, Huang:2025ljp}
applying confinement physics to improve SM measurements and BSM searches.
We look forward to their future experimental realization
that may help solve the riddle of first-generation Yukawa couplings.

%%%%%%%%%%%%%%%%%%%%%%%%%%%%%%%%%%%%%%%%%%%%%%%%%%%%%%%%%%%%%%%%%%%%%%%%%%%%%%%%
\begin{acknowledgments}
\headingAcknowledgments
The author gratefully acknowledges encouraging discussions
with Ankita Budhraja, Eric Laenen, Juraj Klari\'c, Philipp Klose,
Rebecca von Kuk, Ian Moult, Piet Mulders, Tristan du Pree, Zhiquan Sun, and Wouter Waalewijn,
and would like to thank Gavin Salam for an inspiring Nikhef Colloquium in March 2024
emphasizing the mystery of the first-generation Yukawa couplings,
which led the author to doodle \fig{other_processes}~(b) on the margin of his notes from that talk.

The author wishes to thank the CERN Theoretical Physics Department
for hospitality while part of this work was carried out.
The author was supported by the D-ITP consortium, a program of NWO that is funded by the Dutch Ministry of Education, Culture and Science (OCW).
\end{acknowledgments}
%%%%%%%%%%%%%%%%%%%%%%%%%%%%%%%%%%%%%%%%%%%%%%%%%%%%%%%%%%%%%%%%%%%%%%%%%%%%%%%%

\bibliography{references}

\onecolumngrid

%%%%%%%%%%%%%%%%%%%%%%%%%%%%%%%%%%%%%%%%%%%%%%%%%%%%%%%%%%%%%%%%%%%%%%%%%%%%%%%%
\section*{Supplemental material}
\label{supplement}
%%%%%%%%%%%%%%%%%%%%%%%%%%%%%%%%%%%%%%%%%%%%%%%%%%%%%%%%%%%%%%%%%%%%%%%%%%%%%%%%

% shift all counters by 1000 for uniqueness
\setcounter{equation}{1000}
\setcounter{figure}{1000}
\setcounter{table}{1000}

% change formatting of counters to have an "S" for supplement in front
\renewcommand{\theequation}{S\the\numexpr\value{equation}-1000\relax}
\renewcommand{\thefigure}{S\the\numexpr\value{figure}-1000\relax}
\renewcommand{\thetable}{S\the\numexpr\value{table}-1000\relax}

In this supplemental material we collect explicit results for all
asymmetry coefficients and background cross sections.
We also list the estimated degrees of transverse polarization,
describe explicitly how the projected theory systematics are estimated,
and provide the final $\chi^2$ distributions of our sensitivity analysis in numerical form.
This material is not required in any way to follow the presentation in the main text,
but is provided purely for the sake of reproducibility and the reference of the reader.

%===============================================================================
\subsection{Integrated signal cross sections}
\label{sec:integrated_cross_sections}
%===============================================================================

Using \eq{modelling_assumption}, the integrated on-shell signal cross section
in \eq{fact_Delta_sigma_hV} simply becomes
%%%
\begin{align} \label{eq:simplify_Delta_sigma_hV}
\Delta \sigma_{V\!Hh}
&= \frac{2}{\pi} \sum_q \frac{y_q}{e} \Bigl(
   P_{qh} \, \Delta \sigma_{q,V\!H}
   - P_{\bq h} \, \Delta \sigma_{\bq,V\!H}
\Bigr)
\,,\end{align}
%%%
where the coefficients $\Delta \sigma_{q,V\!H}$ and $\Delta \sigma_{\bq, V\!H}$
are defined as
%%%
\begin{align} \label{eq:def_Delta_sigma_qV}
\Delta \sigma_{q,V\!H}
\equiv \frac{e^4}{16 \pi \Ecm^2} \int \! \df \Phi \, \sum_{q'}
C_{qq',V\! H}^{TU} \, f_{q}(x_a) \, f_{\bar{q}'}(x_b)
\,, \qquad
\Delta \sigma_{\bar{q},V\!H}
\equiv \frac{e^4}{16\pi \Ecm^2} \int \! \df \Phi \, \sum_{q'}
C_{q'q,V\!H}^{TU} \, f_{\bar{q}}(x_a) \, f_{q'}(x_b)
\,.\end{align}
%%%
Here $\Ecm$ is the hadronic center-of-mass energy,
$\df \Phi \equiv \df Y_V \, \df Y_H \, \df p_T^V$
is the hard production phase space,
and the PDFs are evaluated at fractions $x_{a,b}$
of the initial-state proton momenta carried by the $VH$ final state.
These coefficient cross sections are readily evaluated
by performing a numerical integral over phase space,
with results collected in \tab{integrated_cross_sections}.
For reference, we also include numerical results
for the total unpolarized $VH$ cross section,
which to leading order in perturbation theory takes the form
%%%
\begin{align}
\sigma_{V\!H} = \frac{e^4}{16\pi \Ecm^2} \int \! \df \Phi \, \sum_{q, q'}
   C^{UU}_{qq'VH} \, \Bigl[
      f_{q}(x_a) \, f_{\bar{q}'}(x_b)
      + f_{\bar{q}'}(x_a) \, f_{q}(x_b)
   \Bigr]
\,.\end{align}
%%%
We use the \texttt{PDF4LHC15\_nnlo\_100} PDF set~\cite{Butterworth:2015oua}
and set the factorization scale to $\mu_F = Q$.
Our electroweak inputs follow \refcite{Tanabashi:2018oca, Ebert:2020dfc}.
We use the complete CKM matrix including offdiagonal entries.
The relevant tree-level hard coefficients are given by
%%%
\begin{align} \label{eq:results_cuu_ctu}
C_{qq',V\! H}^{UU}
&= c_H^2 (c_L^2 + c_R^2) \, \frac{p_T^V}{2 N_c} \frac{
    M_V^2 Q^2 + 4 (p_a \cdot p_V)(p_b \cdot p_V)
}{
   M_V^2 Q^2 ( Q^2 - M_V^2 )^2
}
\,, \nn \\
C_{qq',V\! H}^{TU}
&= - c_H (c_L^2 - c_R^2) \, \frac{(p_{T}^V)^2}{\sqrt{2} N_c } \frac{
   M_V^2 + 2 p_b \cdot p_V
}{
   M_V^2 (p_V - p_b)^2
   (
      Q^2 - M_V^2
   )
}
\,,\end{align}
%%%
where $p_a$ ($p_b$, $p_V$)
is the momentum of the parton incoming along the direction of $h$
(the second incoming parton, the vector boson),
$M_V$ is the mass of $V$,
$c_H \propto M_V$ is the coupling of $V$ to the Higgs boson
and $c_{L,R}$ are the couplings of $V$
to the currents $\bar{\psi}_{\bq'} \, \gamma^\mu P_{L,R} \, \psi_{q}$ in units of $\abs{e}$.
To numerically assemble the signal prediction,
one also requires --- in addition to the estimated DOPs quoted below
--- the values of the quark Yukawa couplings at the reference scale $\mu = m_H$.
These are obtained from the \MSbar{} quark masses at the input scale~\cite{Tanabashi:2018oca},
using for simplicity the leading-logarithmic renormalization group evolution
of the quark Yukawa couplings and the strong coupling.
The numerical values are given in \tab{yukawa_couplings}.
We recall that we have scaled out the DOPs
and the Yukawa couplings from the $\Delta \sigma_{q,V\!H}$,
and that at the same time the total semi-inclusive $VHh$
production cross section
is still larger than $\sigma_{V\!H}$ by a factor of $\exv{n_h}$,
so there is nothing unphysical about some individual asymmetry coefficients $\Delta \sigma_{q,V\!H}$ being larger than the total $\sigma_{V\!H}$.

\begin{table}
\renewcommand{\arraystretch}{1.5}
\tabcolsep 5pt
\centering
\begin{tabular}{cccc}
\hline\hline
$V$ &  $Z$ & $W^+$ & $W^-$ \\
\hline
$\sigma_{V\!H}~[\!\fb]$ & 744.05 & 827.09 & 532.49 \\
\hline
$\Delta \sigma_{d,V\!H}~[\!\fb]$ & 416.42 & 0      & 899.39 \\
$\Delta \sigma_{u,V\!H}~[\!\fb]$ & 375.45 & 1566.9 & 0      \\
$\Delta \sigma_{s,V\!H}~[\!\fb]$ & 95.497 & 0      & 179.50 \\
$\Delta \sigma_{c,V\!H}~[\!\fb]$ & 30.463 & 169.64 & 0      \\
$\Delta \sigma_{b,V\!H}~[\!\fb]$ & 23.395 & 0      & 0.1516 \\
\hline
$\Delta \sigma_{\bd,V\!H}~[\!\fb]$ & 416.42 & 1516.3 & 0      \\
$\Delta \sigma_{\bu,V\!H}~[\!\fb]$ & 375.45 & 0      & 883.43 \\
$\Delta \sigma_{\bs,V\!H}~[\!\fb]$ & 95.497 & 221.65 & 0      \\
$\Delta \sigma_{\bc,V\!H}~[\!\fb]$ & 30.463 & 0      & 194.94 \\
$\Delta \sigma_{\bb,V\!H}~[\!\fb]$ & 23.395 & 0.1591 & 0      \\
\hline\hline
\end{tabular}
\caption{%
Integrated unpolarized cross sections
and asymmetry coefficient cross sections
for $VH$ production at $\Ecm = 14 \TeV$.
}
\label{tab:integrated_cross_sections}
\end{table}

\begin{table}[t]
\renewcommand{\arraystretch}{1.5}
\tabcolsep 5pt
\centering
\begin{tabular}{cccccc}
\hline\hline
$q$ &  $\overline{m}_q(\mu_0)$ & $\mu_0$ & $y_q(\mu_0)$ & $y_q \equiv y_q(\mu = m_H)$ & $\alphas\bigl[\overline{m}_q(\overline{m}_q)\bigr]$ \\
\hline
$d$ & $4.67 \MeV $ & $ 2 \GeV   $      & $2.68231 \times 10^{-5} $ & $ 1.72783 \times 10^{-5} $ & $-$       \\
$u$ & $2.16 \MeV $ & $ 2 \GeV   $      & $1.24064 \times 10^{-5} $ & $ 7.99169 \times 10^{-6} $ & $-$       \\
$s$ & $93.4 \MeV $ & $ 2 \GeV   $      & $5.36462 \times 10^{-4} $ & $ 3.45567 \times 10^{-4} $ & $-$       \\
$c$ & $1.27 \GeV $ & $ \overline{m}_c$ & $7.29451 \times 10^{-3} $ & $ 4.32926 \times 10^{-3} $ & $0.30678$ \\
$b$ & $4.18 \GeV $ & $ \overline{m}_b$ & $2.40087 \times 10^{-2} $ & $ 1.72720 \times 10^{-2} $ & $0.21217$ \\
\hline\hline
\end{tabular}
\caption{%
Input \MSbar{} quark masses, input scales, and Yukawa couplings at the input scales ($\mu_0$)
and the reference scale $\mu = m_H = 125.09 \GeV$, as used for the numerical analysis
in the main text.
For reference, we also provide the value of the LL running coupling
at the scale of the heavy quark mass for $q = c, b$.
}
\label{tab:yukawa_couplings}
\end{table}

%===============================================================================
\subsection{Background cross sections, signal efficiencies, and pile-up}
\label{sec:background_cross_sections_and_efficiencies}
%===============================================================================

\begin{table}
\renewcommand{\arraystretch}{1.5}
\tabcolsep 5pt
\centering
\begin{tabular}{cccc}
\hline\hline
$V$ &  $Z$ & $W^+$ & $W^-$ \\
\hline
$\sigma_{X}~[\!\fb]$ & 116.8 & 85.3 & 66.1 \\
$A^{V\!H}_\mathrm{fid}$ & 0.361 & 0.459 & 0.436 \\
$\eps_{X/V\!H}$ & $6.7 \times 10^{-3}$ & $27.6 \times 10^{-3}$ & $26.2 \times 10^{-3}$ \\
\hline\hline
\end{tabular}
\caption{%
Hard background cross sections $\sigma_X$
from continuum $Vjj$ production and leptonic $V$ decay
(including flavor tagging and fiducial acceptance cuts),
fiducial acceptance factors $A^{V\!H}_\mathrm{fid}$
applied to the $VH$ asymmetry signal,
and combined decay and efficiency factors $\eps_{X/V\!H}$
for the respective channel.
Results are combined across electron ($\ell = e$) and muon channels ($\ell = \mu$).
}
\label{tab:background_cross_sections_and_efficiencies}
\end{table}

We consider on-shell $VH$ production
with the Higgs boson decaying into a $b\bar{b}$ pair
and the vector boson decaying leptonically.
We always combine the respective electron and muon channels (but exclude tauons).
To compute $\sigma_X$ in \eq{delta_stat_delta_sigma_VHh}
we consider the dominant $V j j$ continuum QCD background.
We use \texttt{Sherpa~3.1} to calculate
separate truth-level $Vjj$ cross sections
for all combinations of light jets, $c$ jets, and $b$ jets
at leading order in the strong coupling.
We then apply the following flavor tagging efficiencies
to each of the two jets individually,
assuming the loose flavor-tagging working point
with $70 \%$ $b$ signal efficiency from \refscite{ATLAS:2022qxm, ATLAS:2024yzu},
%%%
\begin{align}
\eps_{b} = 70 \%
\,, \qquad
\eps_{c} = 7.9 \%
\,, \qquad
\eps_\mathrm{light} = 0.18 \%
\,.\end{align}
%%%
These in turn indicate the probability for a truth $b$, $c$, or light jet
to be kept as a $b$ jet.
The resulting total cross section is dominated by the contribution
with two truth-level $b$ jets.
We apply fiducial acceptance cuts of $p_T \geq 25 \GeV$ and $\abs{\eta} > 2.5$
to all jets and charged leptons.
In the case of $Z \to \ell^+ \ell^-$ we also cut
on the dilepton invariant mass, $66 \leq m_{\ell\ell} \leq 116 \GeV$.
To account for the effect of finite jet energy resolution in a simplified way
(and more generally, as a simplified proxy
for multivariant background discrimination
with the dijet invariant mass as the dominant input),
we cut on $105 \leq m_{jj} \leq 145 \GeV$
and assume that this retains all of the Higgs signal.
The resulting hard background production cross sections $\sigma_X$
are given in \tab{background_cross_sections_and_efficiencies}.
We have verified that with these simplified assumptions,
the expected statistical uncertainties
on the unpolarized $VH(\to b \bar{b})$ signal strengths
at $\cL = 140 \fb^{-1}$ are comparable to those obtained in \refcite{ATLAS:2024yzu}.

The decay and reconstruction efficiency factor $\eps_{X/V\! H}$
applied to the on-shell $VH$ asymmetry signal is given by
%%%
\begin{align}
\eps_{X / V\! H}
= \eps_b^2 \,
A_\mathrm{fid}^{V\!H} \,
\mathrm{Br}(V \to \text{leptons}) \,
\mathrm{Br}(H \to b\bar{b})
\,,\end{align}
%%%
where $\eps_b^2$ is the $b$ tagging efficiency as above
and $A_\mathrm{fid}^{V\!H}$ is the acceptance factor of the fiducial cuts
on the $V$ and $H$ decay products.
For the relevant branching ratios we use $\mathrm{Br}(H \to b\bar{b}) = 58.09 \%$~\cite{LHCHiggsCrossSectionWorkingGroup:2016ypw},
while for $\mathrm{Br}(V \to \text{leptons})$
we take twice the value of
the branching ratio to a given fixed lepton flavor,
$\mathrm{Br}(Z \to \ell^+ \ell^-) = 3.3658 \%$
and $\mathrm{Br}(W \to \ell \nu) = 10.86 \%$~\cite{Tanabashi:2018oca},
since we combine electrons and muons.
For simplicity, we obtain the acceptance factors $A_\mathrm{fid}^{V\!H}$
from \texttt{Sherpa} simulations of the unpolarized $VH$ signal.
The accuracy of this approximation is on par with our modelling assumption
in \eq{modelling_assumption}, and can easily be refined in the future.
Results for $A_\mathrm{fid}^{V\!H}$
and the final $\eps_{X / V\! H}$ applied to the on-shell results
in \tab{integrated_cross_sections}
are reported in \tab{background_cross_sections_and_efficiencies}.

A final component of the statistical uncertainty estimate
in \eq{delta_stat_delta_sigma_VHh} are the effects
of pile-up, which tends to populate in particular the forward detector region
with hadronic activity,
and of pile-up suppression techniques.
Only tracks that at truth level originate from the hard vertex
carry the asymmetry, while pile-up tracks are isotropic
and dilute the statistical power.
We assume that dedicated forward silicon tracker coverage at $\abs{\eta} > 2.5$
as e.g.\ described in \refcite{CERN-LHCC-2017-021, Rossi:2022dcv}
will have become available to reconstruct forward charged-particle tracks
(i.e., mostly charged pions)
and associate them with the hard vertex,
possibly aided by forward timing information~\cite{CERN-LHCC-2020-007}.
The effect of pile-up and pile-up suppression
is quantified by the signal efficiency
$\eps_h$ of the pile-up suppression cuts
and the signal purity $P_\mathrm{PU}^h$.
We define $P_\mathrm{PU}^h$ as the fraction of truth hard tracks
within the total sample of tracks that
are associated with the hard vertex by the vertex finding method,
while $1 - P_\mathrm{PU}^h$ is the residual fraction of truth pile-up tracks.
The expected total number of forward hadrons
reconstructed in association with $X$,
which sets \eq{delta_stat_delta_sigma_VHh},
is then given by
%%%
\begin{align}
2 N^\pm_{Xh} \approx N^+_{Xh} + N^-_{Xh} = \sigma_{Xh} \, \cL
\,, \qquad
\sigma_{Xh} = \frac{\eps_h \, \exv{n_h} \, \sigma_X}{P_\mathrm{PU}^h}
\,.\end{align}
%%%
In \refcite{CERN-LHCC-2017-021}, the performance
of the primary vertex reconstruction was assessed
for a detector region similar to our fiducial forward region of interest
and for realistic HL-LHC conditions
with $\exv{\mu} = 200$ pile-up events per hard scattering.
Remarkably, the extended tracker coverage
is projected to achieve purities of $50 \! - \! 70 \%$
for tracks with $2.5 \leq \abs{\eta} \leq 4$
and $p_T \geq 1 \GeV$ (see Fig.~3.18 in \refcite{CERN-LHCC-2017-021}, top left)
at signal efficiencies of $20 \! - \! 40 \%$ (top right).
For definiteness we take $P_\mathrm{PU}^h = 50\%$ and $\eps_h = 30 \%$
for all hadron species we consider, where we assume for simplicity
that the pile-up rejection for forward kaons and $D$ mesons
reconstructed through their decays
is comparable to the case of plain charged-pion tracks.

%===============================================================================
\subsection{Estimated degrees of transverse polarization and theory systematics}
\label{sec:dops}
%===============================================================================

For the light/light DOPs, we have two doubly-favored flavor combinations
$P_{d \pi^+} = P_{u \pi^-} = P_{\fav^2}$
where the ``quark hole'' has the quantum numbers of the pion valence antiquark
and the pion valence quark is already present in the proton.
For these cases we take $P_{\fav^2} = 1$.
We stress that even for an estimate of $P_{qh} = 1$ in our normalization here
(i.e., relative to the PDF),
the ratio $\bar{h}^\perp_{qh}/\bar{f}_{qh} \sim P_{qh}/\exv{n_h} \approx 0.1$
to the unpolarized FrF
is far from its actual positivity bound $\leq 1$,
and in line with $\lesssim f_\pi/m_p$ as expected from chiral symmetry breaking.
(It is also in line with the observed size of the valence dihadron IFF
at large momentum fractions of the dihadron pair~\cite{Courtoy:2012ry, Cocuzza:2023vqs}.)
We further have two singly-favored cases $P_{\bu \pi^+} = P_{\bd \pi^-} = P_\fav = 0.5$,
where only the ``quark hole'' quantum numbers agree,
but an antiquark is required from the remnant.
Other combinations of $u$ and $d$ (anti)quarks with $\pi^\pm$ are disfavored, $P_\dis = 0.1$.

For heavier quarks $Q = s, c, b$, it is useful to separately consider the total (symmetrized)
DOP $\bar{P}_{Q h} \equiv (P_{Q h} + P_{\bQ h})/2$ and the sea-quark asymmetry
$\Delta P_{Q h} \equiv (P_{Q h} - P_{\bQ h})/2$.
For the strange/pion DOP we take $\bar{P}_{s h} = P_\dis (1 \pm 3)$
and, to be conservative, the extreme case $\Delta P_{s h} = (0 \pm 1) \bar{P}_{s h}$.
To estimate the heavy/pion DOPs with $Q = c,b$, we work to the leading nontrivial power in $\lqcd \ll m_Q$
and find, using techniques from \refcite{vonKuk:2023jfd},
%%%
\begin{align} \label{eq:dops_heavy_pion}
\bar{P}_{Q\pi^\pm} = \frac{\Lambda_{g\pi^\pm}}{m_Q}(1 \pm 3)
\,, \qquad
\Delta P_{Q\pi^\pm} = \frac{\as(m_Q)}{\pi} \frac{\Lambda_{g\pi^\pm}}{m_Q}(0 \pm 5)
\,,\end{align}
%%%
where formally the size of $\Lambda_{g\pi^\pm} = 0.3 \GeV$ is set by the integral of a twist-3 gluon/pion FrF.
We note that while the matching onto this twist-3 FrF
will involve an additional power of $\alphas(m_Q)/\pi$,
the latter drops out since the DOPs are defined relative to $f_Q \sim (\as/\pi) \, f_g$.
The estimate for the sea asymmetry follows by noting
that perturbatively, heavy quarks and antiquarks
only become distinguished by higher-order color structures like $d^{abc} \, d^{abc}$.
The parametric uncertainties on the above estimates give rise to the theory covariance matrix
used to construct the $\chi^2$ function in the main text.
We note that while at this stage the uncertainties are uncorrelated,
since these are estimates of independent nonperturbative matrix elements,
they do of course induce correlations between all vector boson and hadron channels
that the parameters appear in.

For the case of reconstructed kaons and $D$ mesons,
we simply set the first-generation DOPs to zero,
$P_{q H} = P_{\bq H} = 0$,
where $H = K^\pm, D^\pm$ and $q = u, d$.
The strange/kaon DOPs are
$P_{s K^+} = P_{\fav^2}$,
$P_{\bs K^-} = P_\fav$,
and $P_{s K^-} = P_{\bs K^+} = P_\dis$.
For the heavy/kaon DOPs and sea asymmetries involving $Q = c,b$
we again use \eq{dops_heavy_pion} with $\Lambda_{gK^\pm} = 0.3 \GeV$ as its central value.
For signal charm mesons $D^\pm$, estimates for all BMFrFs
are essentially perturbative~\cite{vonKuk:2023jfd}.
At leading power in $\lqcd/m_c$
they only involve the twist-2 gluon PDF $f_g \sim f_{u,d}$
and an $\ord{1}$ coefficient $\chi_{D^\pm}$ encoding the probability for a free $c$ (anti)quark to fragment
into the experimentally selected set of $D$ meson states, where we take
%%%
\begin{alignat}{3}
P_{\bc D^+}
&= P_{c D^-}
= \frac{\as(m_c)}{\pi}
\,, \qquad
P_{c D^+} &&= P_{\bc D^-} = P_{s D^+} = P_{s D^-} = \frac{\as^2(m_c)}{\pi^2}
\,, \nn \\
\bar{P}_{b D^\pm} &= \frac{\as^2(m_b)}{\pi^2}(1 \pm 3)
\,, \qquad
\Delta P_{b D^\pm} &&= \frac{\as^3(m_b)}{\pi^3}(0 \pm 5)
\,.\end{alignat}
%%%
Here the leading contribution to the valence BMFrF starts at $\as^2/\pi^2$
(or $\as/\pi$, relative to $f_c$) and arises from absorptive contributions
involving an additional Wilson-line attachment~\cite{vonKuk:2023jfd}.

The second main source of systematic theory uncertainties on the measurement of $y_q$ from $\Delta \sigma_{V\!Hh}$
are subleading terms in the underlying expansion in $\lqcd, m_q, p_T^h \ll Q$
that allowed us to make use of leading-twist factorization in \eq{fact_Delta_sigma_hV}.
The leading contribution of this kind again arises from twist-3 collinear FrFs~\cite{Chen:2023wsi},
in this case convolved with a hard matching coefficient for the $VH$ signal process.
(We refer to \refcite{paper2} for details.)
We numerically estimate the impact of these terms as
%%%
\begin{align} \label{eq:delta_theo_tw3_delta_sigma_VHh}
\Delta \sigma_{V\!Hh}^\mathrm{tw3}
= \frac{1 \GeV}{Q_\mathrm{ref}} \, \Delta A_\mathrm{tw3} \, \sigma_{V\!H} \, (1 \pm 0.5)
\times \begin{cases}
    \exv{n_h} \,, &h = \pi^\pm, K^\pm \,,
    \\
    10 \, \exv{n_h} \,, &h = D^\pm
\,,\end{cases}
\end{align}
%%%
where the numerator of $1 \GeV$ is a typical value of $p_T^h \gtrsim \lqcd$
(which would dominate the twist expansion if one were to admit $p_T^h \gg \lqcd$)
and we use $Q_\mathrm{ref} = 250 \GeV$ as a representative value for the hard scale.
For light hadrons we have included a factor of $\exv{n_h}$ to reflect the fact that
the underlying twist-3 FrFs should be suppressed by $\lqcd$ (or $p_T^h$)
relative to the unpolarized FrFs, not the PDFs that appear in the unpolarized $VH$ signal cross section $\sigma_{V\!H}$.
For $D$ hadrons, where $\exv{n_h} \sim \as(m_c)/\pi \ll 1$, this would be too aggressive,
since further study is required to determine
what properties of the twist-2 matrix elements carry over to twist-3 heavy-quark FrFs.
We therefore include an additional factor of $10$ in this case.
The additional factor $\Delta A_\mathrm{tw3}$ multiplying $\sigma_{V\!H}$ accounts for the fact
that the twist-3 terms for $VH$ production
are odd in $\Delta Y \equiv Y_Z - Y_H$~\cite{paper2}, unlike the asymmetry signal coefficient in \eq{results_cuu_ctu},
and thus contribute only indirectly through fiducial acceptance cuts, similar to the $A_i$ angular coefficients in Drell-Yan~\cite{Ebert:2020dfc}.
As a rough estimate of this residual effect we take $\Delta A_\mathrm{tw3} = 0.1$.
We also assume that like the BMFrF, the dominant favored and doubly-favored twist-3 FrFs
can be extracted to a reasonable accuracy ($\sim 50\%$)
from baseline data on Drell-Yan hadron-hadron correlations~\cite{paper2}.
The above estimate of the overall size of twist-3 effects
is conservative when compared to e.g.\ the observed size
of the Qiu-Sterman function (see, for example, \refcite{Echevarria:2020hpy} for a recent determination)
which in units of the unpolarized PDF is found to be closer to $N_q \sim 100 \MeV \ll m_\pi \sim 1 \GeV$.

%===============================================================================
\subsection{Combined \texorpdfstring{$\chi^2$}{chi2} statistics}
\label{sec:chi2_statistics}
%===============================================================================

For a given pair of flavors $(q, q')$, specifically $(u, d)$
for the left panel and $(c, s)$
for the right panel of \fig{projected_limits},
we define the shorthand
$\Delta \vec{\kappa} \equiv (\Delta \kappa_q, \Delta \kappa_{q'})^T$.
Then, since all our constraints are linear,
the combined $\chi^2$ distributions shown
in \fig{projected_limits}
take the form
$\chi^2 = \Delta \vec{\kappa} \cdot \tensor{A} \cdot \Delta \vec{\kappa}$,
with a vanishing $\chi^2 = 0$ for the best fit at $\Delta \vec{\kappa} = 0$
by construction.
The parameter covariance matrices $\tensor{A}$
for the final combined limits are numerically given by
%%%
\begin{align}
\tensor{A}^{(u,d)}_\stat
= \begin{pmatrix}
 6.2812 \times 10^{-7} & -8.7716 \times 10^{-7} \\
-8.7716 \times 10^{-7} &  1.8677 \times 10^{-6}
\end{pmatrix}
\,, \qquad
\tensor{A}^{(u,d)}_{\syst}
= \begin{pmatrix}
 0.00028721 & -0.00038608 \\
-0.00038608 & 0.00084488
\end{pmatrix}
\end{align}
%%%
for the light-quark case, while for the heavy-quark case we have
%%%
\begin{align}
\tensor{A}^{(c,s)}_\stat
= \begin{pmatrix}
 1.1458 \times 10^{-2} & -3.9296 \times 10^{-4} \\
-3.9296 \times 10^{-4} &  8.1437 \times 10^{-5}
\end{pmatrix}
\,, \qquad
\tensor{A}^{(c,s)}_{\syst}
= \begin{pmatrix}
 14.653 & -0.15601 \\
 -0.15601 &  0.064700
\end{pmatrix}
\,.\end{align}
%%%

\end{document}